%% file: ms.tex
\documentclass[conference,final]{IEEEtran}
\IEEEoverridecommandlockouts
\usepackage[noadjust]{cite}
\usepackage{amsmath,amssymb,amsfonts}
\usepackage{graphicx}
\usepackage{textcomp}
\usepackage{xcolor}
\include{include}

\renewcommand{\code}[1]{#1}
\renewcommand{\enzo}{ENZO}
\renewcommand{\enzop}{\mbox{Enzo-P}}
\renewcommand{\enzoe}{\mbox{Enzo-E}}
\renewcommand{\cello}{Cello}
\renewcommand{\charm}{Charm++}

\def\BibTeX{{\rm B\kern-.05em{\sc i\kern-.025em b}\kern-.08em
    T\kern-.1667em\lower.7ex\hbox{E}\kern-.125emX}}

\begin{document}

\title{Computational Cosmology and Astrophysics on Adaptive Meshes using Charm++\\
  \thanks{Funding for this work has been provided by the National
    Science Foundation through grants
SI2-SSE-1440709, PHY-1104819, AST-0808184, and OAC-1835402} }

\author{\IEEEauthorblockN{James Bordner}
\IEEEauthorblockA{\textit{San Diego Supercomputer Center} \\
\textit{University of California, San Diego}\\
La Jolla, CA, U.S. \\
jobordner@ucsd.edu}
\and
\IEEEauthorblockN{Michael L.~Norman}
\IEEEauthorblockA{\textit{San Diego Supercomputer Center} \\
\textit{University of California, San Diego}\\
La Jolla, CA, U.S. \\
mlnorman@ucsd.edu}
}

\maketitle


\begin{abstract}
  \todo Astrophysical and cosmological phenomena involve a large variety of
  physical processes, and can encompass an enormous range of scales.
  To effectively investigate these phenomena computationally,
  applications must similarly support modeling these phenomena on
  enormous ranges of scales; furthermore, they must do so efficiently
  on high-performance computing platforms of ever-increasing
  parallelism and complexity.
  We describe Enzo-P, a Petascale redesign of the ENZO adaptive mesh
  refinement astrophysics and cosmology application, along with Cello,
  a reusable and scalable adaptive mesh refinement software framework,
  on which Enzo-P is based.
  Cello's scalability is enabled by the Charm++ Parallel Programming
  System, whose data-driven asynchronous execution model is ideal for
  taking advantage of the available but irregular parallelism in
  adaptive mesh refinement-based applications.
  We present scaling results on the NSF Blue Waters
  supercomputer, and outline our future plans to bring Enzo-P to the
  Exascale Era by targeting highly-heterogeneous accelerator-based
  platforms.
\end{abstract}


\begin{IEEEkeywords}
  Adaptive mesh refinement,
  Astrophysics,
  Octrees,
  Petascale computing
\end{IEEEkeywords}

\section{Introduction}

\begin{figure}[b]
\centerline{\includegraphics[width=3.0in,height=2.18in]{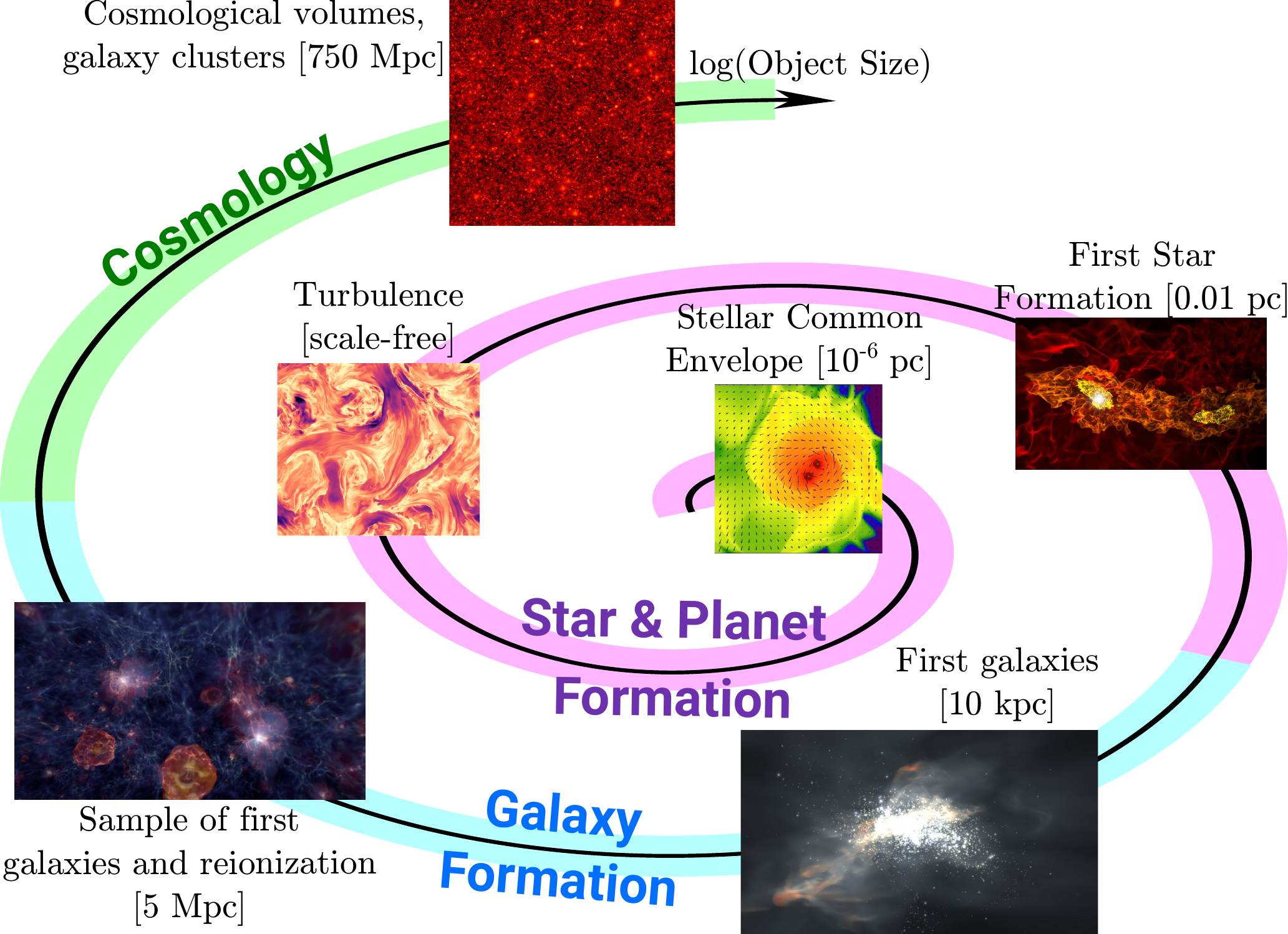}}
\caption{Astrophysical phenomena can range over extreme scales.}
\label{f:scales}
\end{figure}

\todo\pargraph{astro/cosmo application requirements}
There are numerous astrophysics and cosmology topics of scientific
interest, such as early star formation, galaxy formation, galaxy
clusters, and cosmic reionization.  These astrophysical phenomena
typically involve a variety of physical processes, including
hydrodynamics; gravity; gas chemistry, heating, and cooling; radiative
transfer; and cosmological expansion.  Investigating astrophysical
phenomena computationally thus requires a powerful multiphysics
software application with a wide variety of numerical methods.
Astrophysical phenomena also encompass an enormous range of spatial
and temporal scales (see Fig.~\ref{f:scales}).  Astrophysical
simulations must be able to span these scales, sometimes in the same
simulation, such as when simulating early star formation using
cosmological initial conditions~\cite{No08}.  Astrophysics
applications thus require some means to efficiently represent a wide
dynamic range of scales.

\todo\pargraph{ENZO astro/cosmo application}
\textit{ENZO}~\cite{BrNo13} is an MPI parallel application designed
for multiphysics astrophysical and cosmological science simulations,
and was the first such application to use adaptive mesh refinement
(AMR)~\cite{BeOl84,BeCo89} to increase the spatial and temporal
dynamic range~\cite{BrNo97}.
However, while powerful in terms of multiphysics and multiresolution
capabilities, \enzo\ does have some limitations with respect to its
parallel scalability.
In particular, although \enzo\ scales well in uniform grid mode, it
does not scale well beyond a few thousand cores when AMR is used,
limiting its applicability.  This is not
surprising, since development on \enzo\ began in the early 1990's when
``extreme scale'' meant hundreds of CPU's, not today's millions of
cores.
Since then, attempting to improve \enzo's scalability by over $\times 1000$ to
keep pace with the relentless increase in HPC platform parallelism has
become increasingly difficult, and many of \enzo's remaining scaling
bottlenecks cannot be resolved without a fundamental overhaul of its
parallel AMR design and implementation.

This motivated us to develop the ``Petascale \enzo'' fork of the
\enzo\ code called \textit{\enzop}, using a highly scalable AMR
framework called \textit{\cello}~\cite{BoNo12}, which we are
developing concurrently.
Key features of \cello\ are that it implements a fully distributed
``array-of-octrees'' AMR approach, and \cello\ is parallelized using
\charm\ \cite{wwwcharm} rather than MPI.

Below we discuss the \enzop\ / \cello\ / \charm\ software stack in
\S\ref{s:ecc}, we present weak-scaling results of \enzop\ /
\cello\ / \charm\ in \S\ref{s:scaling}, and we conclude with our
future plans in \S\ref{s:future}.

\section{\enzop\ / \cello\ / \charm}
\label{s:ecc}

\todo\pargraph{sections} In this section we describe the software
stack of \enzop\ / \cello\ / \charm\ from the bottom up: in
\S\ref{s:charm} we review
the \charm\ parallel programming system,
in
\S\ref{s:cello} we describe the \cello\ adaptive mesh refinement
framework,
and in
\S\ref{s:enzop} we discuss the \enzop\ science application layer.
%

\subsection{The \charm\ parallel language} \label{s:charm}

   \charm~\cite{KaBh13,wwwcharm}, developed at the Parallel Programming
   Laboratory (PPL) at the University of Illinois, is the visionary
   parallel programming system on which \cello\ is built.
   Since its first public release over $20$ years ago, \charm\ has
   been continuously enhanced and improved by researchers in the PPL
   in collaboration with application developers in diverse areas of
   science and engineering.

   In \charm\ programs, the fundamental parallel object is a
   \textit{chare}.  Chares are C++ objects that contain special
   methods called \textit{entry methods}.  Entry methods may be
   invoked remotely by other chares via \textit{proxies}, and
   communicate with each other using \textit{messages}.  Related
   chares may be grouped together into a \textit{chare array}, in
   which individual chares are accessed using an array proxy plus 
   element index.  Additionally, the \charm\ runtime system supports
   automatic dynamic load balancing of chares within chare arrays.
   The \textit{runtime system} manages chares, assigning their
   location in distributed memory, dynamically migrating chares to
   balance load, communicating message data between chares, and
   dynamically scheduling and executing entry methods.

   Numerous science and engineering applications have been developed
   using \charm.  Perhaps the best known is
   \namd, a Gordon Bell and Sidney Fernbach Award-winning
   parallel molecular dynamics code, which has scaled to beyond $500K$
   cores~\cite{wwwnamd}.
   Other \charm\ codes include \openatom~\cite{wwwopenatom}, a highly
   scalable quantum chemistry application; \changa~\cite{wwwchanga}, a
   scalable collisionless N-body code for cosmological simulations;
   and \episimdemics\cite{wwwepisimdemics}, a simulation system for
   studying the spread of contagion over large interaction networks.

   Applications built on \charm\ directly benefit from its natural
   latency tolerance, overlap of communication and computation,
   dynamic load balancing, and checkpoint / restart capabilities.
   Emerging scalability issues, including fault-tolerance and
   improving energy efficiency, are also provided by \charm, with
   active research in energy-aware rollback-recovery~\cite{MeSa14},
   fault-tolerance protocols~\cite{MeSa12}, automated checkpoint /
   restart mechanisms~\cite{NiMe13}, and automated thermal-aware load
   balancing~\cite{MeAc14}.
   Since \charm\ is based on an introspective and adaptive runtime
   system, it is well-suited to meet the challenges of increasingly
   complex hardware and software, and is poised to be a programming
   model of choice for the Exascale Era.  

\begin{figure}[tb]
  \centerline{
    \includegraphics[width=2.50in]{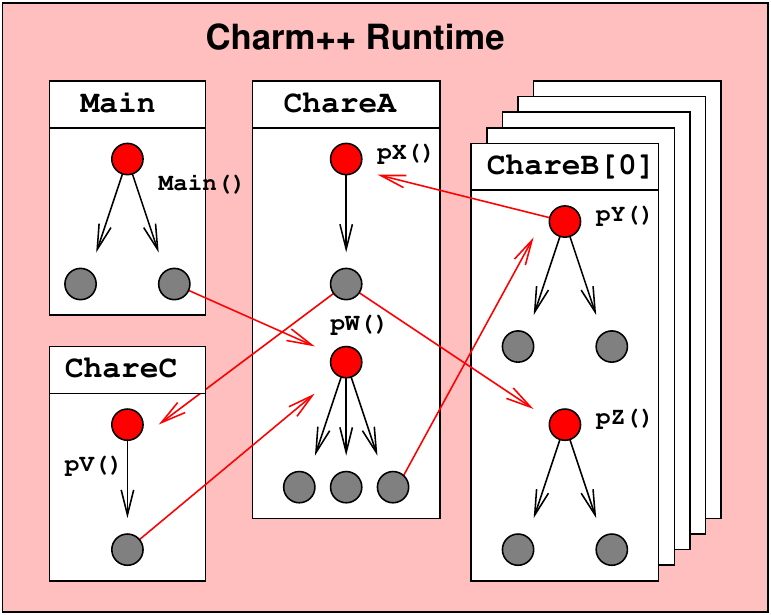}}
  \caption{\charm\ supports object-based data-driven asynchronous parallel programming.}
\label{f:charm}
\end{figure}

\subsection{The \cello\ AMR framework} \label{s:cello}

\begin{figure}[t]
  \centerline{\includegraphics[width=1.5in]{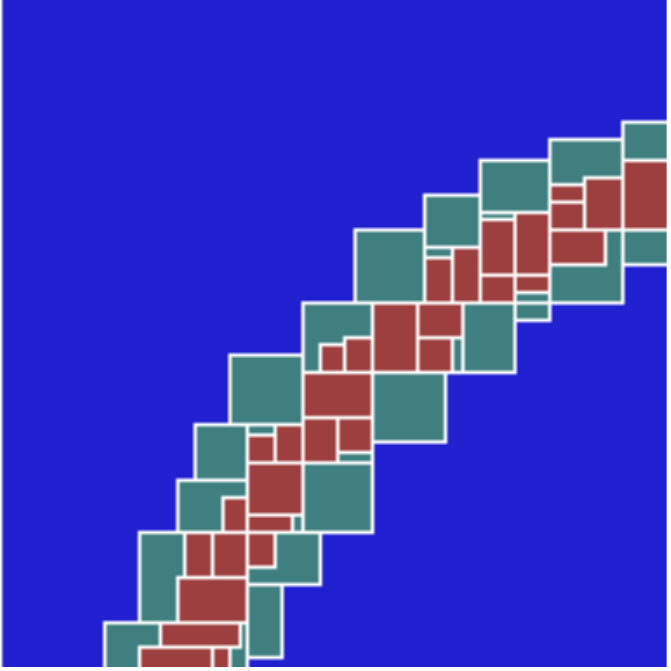}
  \includegraphics[width=1.5in]{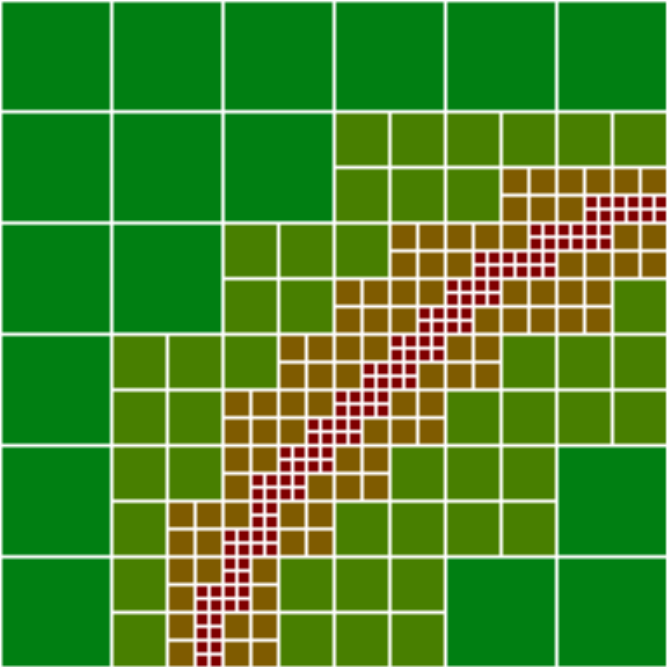}}
  \caption{Whereas \enzo\ uses structured AMR (left), \enzop\ uses ``array-of-octree'' AMR (right) provided provided by \cello.}
\label{f:amr}
\end{figure}

To enable highly scalable multi-resolution simulations in \enzop, we
are developing \cello, an extremely scalable adaptive mesh refinement
framework.  Of the two commonly used AMR approaches, structured AMR
(SAMR) and octree-based AMR, we decided to implement an octree-based
approach in \cello, even though \enzo\ itself uses SAMR (see
Fig.~\ref{f:amr}).  This design decision was made due to its
demonstrated high-scalability~\cite{BuWi10}, its relatively uniform
parallel task sizes, and the relative simplicity of its mesh hierarchy
data structure.  \cello\ implements a
slightly more general ``array-of-octrees'' approach, to allow for
non-cubical computational domains.

   \cello\ uses \charm\ to implement two parallel data
   structures: a ``\code{Simulation}'' process group for storing data
   associated with each process, and a \code{Block} chare array for
   representing the distributed AMR hierarchy, where each \code{Block}
   is associated with a single node in the array-of-octrees.
   Each \code{Block} contains a \code{Data} object, which stores the
   field and particle data associated with that block.  \code{Field}
   and \code{Particle} objects in \cello\ provide applications with
   easy-to-use API's for accessing field and particle data on the
   \code{Block}.

   \cello\ also provides simple C++ base classes for
   \cello\ applications to inherit from to implement computational
   methods (\code{Method}), initial and boundary conditions
   (\code{Initial} and \code{Boundary}), refinement criteria
   (\code{Refine}), inter-level data interpolation or coarsening
   (\code{Restrict} and \code{Prolong}), linear solvers
   (\code{Solver}), I/O (\code{Output}), etc.  Adding new
   functionality to a \cello\ application, say a new physics kernel or
   a new refinement criterion, is straightforward; typically, it
   involves adding a new inherited C++ class and implementing one
   or two virtual methods, which operate on a single \code{Block}.
   This object-oriented approach helps provide extensibility,
   manageability, and composability to \cello\ applications.

\begin{figure}[b]
  \centerline{\includegraphics[width=1.5in]{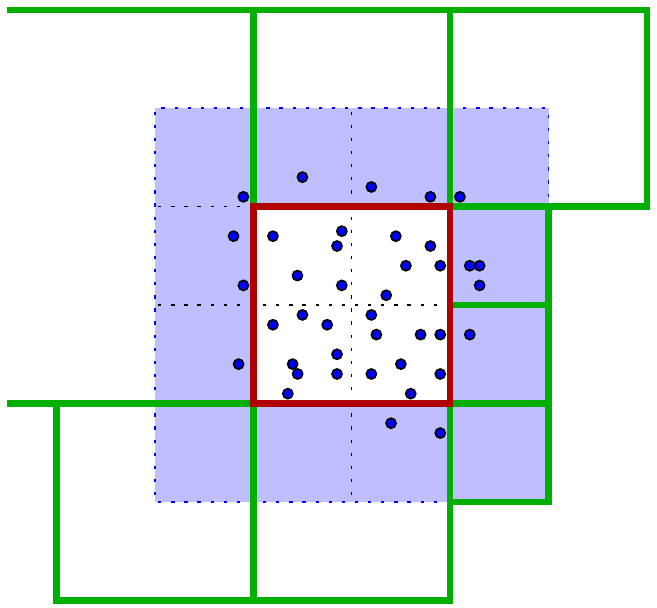}\
    \includegraphics[width=1.5in]{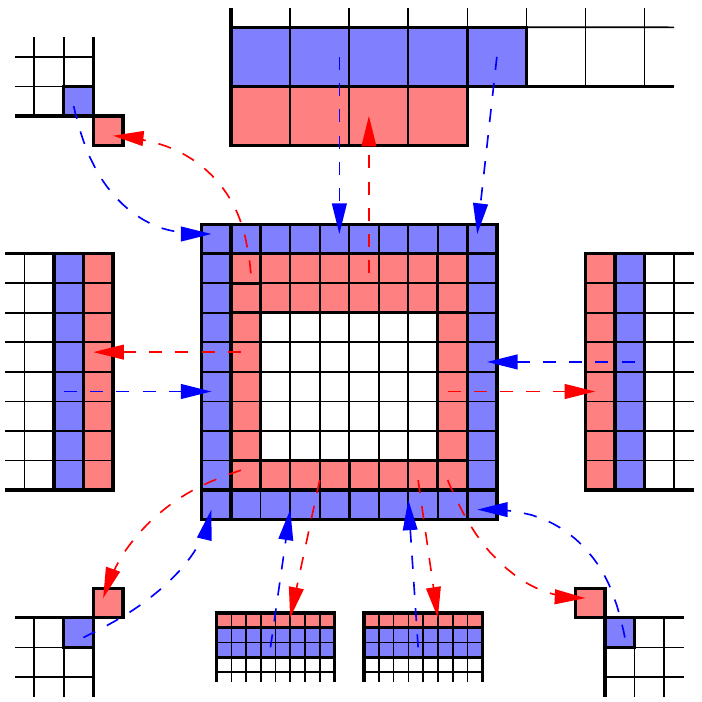}}
  \caption{\cello\ supports both Lagrangian particles and Eulerian
    grid computations.  Particles may move between \code{Block}s
    (left), and grid boundary ``ghost'' data are communicated between
    \code{Block}s asynchronously, with the last update received
    triggering the next \code{Method}.}
\label{f:refresh}
\end{figure}

   User-implemented \code{Method}'s acting on a \code{Block} typically
   require a layer of ``ghost'' data surrounding the block to be
   up-to-date, despite the data being assigned to a neighboring
   \code{Block} (see Fig.~\ref{f:refresh}).  This is handled entirely
   by \cello, with the application only needing to specify which
   \code{Field}'s ghost zones need to be updated, and how many cells
   wide the ghost data layer is.
   
   Data locality and optimizing data movement are increasingly crucial
   for high performance scalable parallel software.  While inter-node
   data locality and movement is controlled by \charm\ through its
   wide variety of leading-edge dynamic load balancing algorithms,
   intra-node data locality is handled by \cello.  \textsf{Field} and
   \textsf{Particle} objects organize data in memory to
   improve cache memory hierarchy performance.  This can be done by
   specifying the order of field and particle attributes, aligning
   field memory addresses in memory, adding extra padding between
   field arrays, allocating particle data in fixed-sized batches to
   reduce memory management overhead, and interleaving field values or
   particle attributes.

\subsection{The \enzop\ astrophysics application} \label{s:enzop}

\todo\pargraph{\enzop\ history and current status}
\enzop\ is the
astrophysics and cosmology application being developed using \cello.
While it does not yet support the full range of physics capabilities
of its parent application \enzo, it has reached the point where it is
capable of running scientifically viable cosmological simulations of
sizes limited only by the available HPC hardware.

\enzop\ supports a growing range of numerical methods,
including the core hydrodynamics and gravity solvers, as well as
chemistry and cooling via the GRACKLE software library \cite{SmBr17}.
The main hydrodynamical solver is a modified piece-wise parabolic
method (PPM)~\cite{WoCo84,BrNo95}, implemented as a
\cello\ \code{Method} (\code{ppm}).  Another PPM method,
PPML~\cite{UsPo09}, is available for magnetized supersonic turbulence
simulations (\code{ppml}).
\enzop\ currently solves for the gravitational potential using a
cloud-in-cell particle-mesh (PM) method (\code{gravity}).  The
\code{Method} can one of several linear solvers, including
preconditioned CG and BiCG-STAB Krylov solvers (\code{cg} and
\code{bicgstab}), or a geometric multigrid V-cycle solver
(\code{mg0}).  Our current linear solver of choice is
HG, developed by Dan Reynolds, and implemented as a composite of
other available \code{Solvers}.
\enzop\ also supports cosmological units, reading cosmological initial
conditions from HDF5 files, and cosmological comoving expansion terms.

\section{Parallel scaling results} \label{s:scaling}

Demonstrative weak scaling tests on the NSF Blue Waters (BW) Petascale
platform indicate that \enzop\ AMR hydrodynamics and (non-AMR)
cosmology simulations can scale extremely well in both time and
memory.  (AMR cosmology scaling tests are underway, and results are
expected by the time this paper is available.)  All tests were run
using \charm\ configured for SMP-mode, and to use the native Cray GNI
network layer.

\begin{figure}[tb]
  \centerline{\includegraphics[width=3.0in]{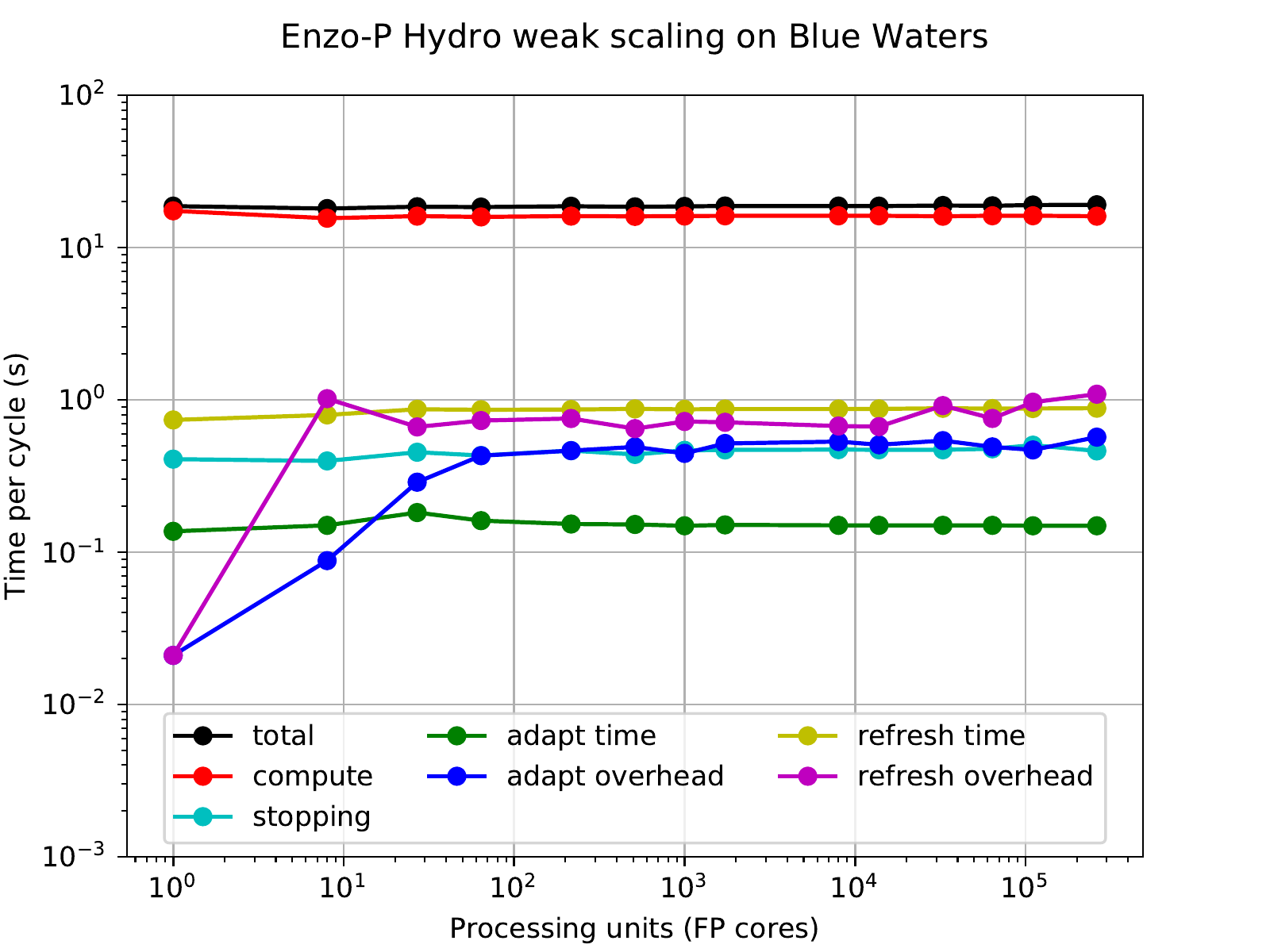}}
  \caption{Weak scaling of \enzop\ / \cello\ hydrodynamics on up to
    $262$K floating-point cores of Blue Waters}
\label{f:scaling:hydro}
\end{figure}

Fig.~\ref{f:scaling:hydro} shows the weak scaling results of a
hydrodynamic test problem with AMR and tracer particles. The problem
involves a 3D array of blast waves, with one blast wave per processor
core. The weak scaling test involves varying the array size from $1^3$
to $64^3$. To inhibit lockstep mesh refinement, instead of a sphere,
the blast is sourced by a high pressure region in the shape of a
letter of the alphabet chosen at random for each core. The initial
state for each blast problem is refined by a 5-level octree, which
resulted in on average about 200 \code{Blocks} (chares) per octree
(core), (such over-decomposition is key to \charm's high efficiency).
Here, the \code{Block}'s grid size is $32^3$ cells, so each blast wave
is initialized with about 6.5M cells and particles. The largest
problem run had $64^3=262,144$ cores (octrees), 52.7M chares, and 1.4
trillion cells plus particles.  We note that this demonstration test
problem, despite its relative simplicity and balanced workload, is
well beyond \enzo's ability to run---it would require about $182$GB
per process simply to store the AMR metadata!

As shown in Fig.~\ref{f:scaling:hydro}, the AMR overhead remains a
small fraction of the total cost through to the largest problem run.
The parallel efficiency is very good, achieving about 94\% at $262$K
floating-point cores. Memory scaling is virtually ideal, due to the
chare array storing the array-of-octrees data being fully distributed
(i.e., no replicated data).

\begin{figure}[b]
  \centerline{\includegraphics[width=3.0in]{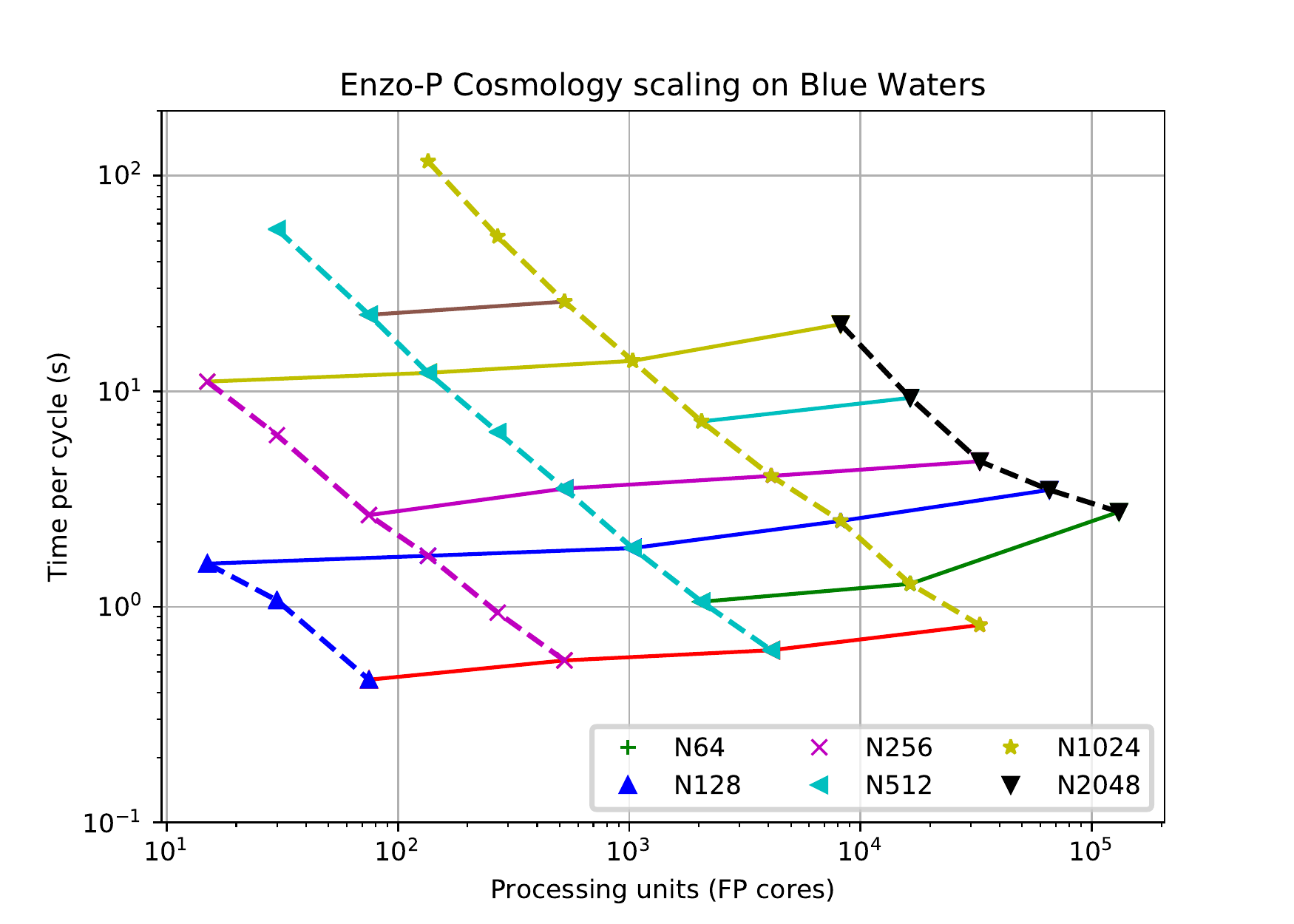}}
  \caption{Weak (dashed line) and strong (solid line) scaling of a basic
    cosmology simulation (uniform grid with hydrodynamics plus dark
    matter) on up to $128$K cores.}
\label{f:scaling:cosmo}
\end{figure}

Fig. \ref{f:scaling:cosmo} shows scaling results for a simple
cosmology simulation on a uniform grid. The problem solves the coupled
equations for single species adiabatic gas dynamics, dark matter
N-body dynamics, and self-gravity on a uniform mesh seeded with
cosmological perturbations. Here a chare is a $32^3$ block containing
hydrodynamic fields and dark matter particles. Self-gravity is
computed using \enzop's \code{mg0} multigrid V-cycle solver.  Scaling
tests consist of grid sizes ranging from $64^3$ to $2048^3$
cells/particles, fixing the number of chares per core. We see
excellent weak scaling results (solid lines), with deviations from
ideal reflecting the $O(\log P)$ scaling of the multigrid algorithm
itself, and some tailoff due to work starvation at the highest core
count (128K) for the relatively small $2048^3$ problem.  Strong
scaling results are excellent as well (dashed lines), and again show
some tailoff due to work starvation.


\section{Future work} \label{s:future}

\enzop\ has demonstrated excellent scaling on Petascale
platforms, as enabled by \charm's data-driven asynchronous approach.
As we continue augmenting the physics capabilities of \enzop\ to that
of its parent code \enzo, we are also preparing for astrophysics and
cosmology at the Exascale.

While \charm\ is specifically targeting supporting Exascale
applications, achieving the required strong scalability on highly
heterogeneous architectures will likely require comprehensive
rethinking of software at all levels, including the \cello\ AMR
software framework and even the application layer itself.

Our approach to developing \enzoe\ will be to collaborate with the
\charm\ group, which has already implemented techniques to simplify
programming of heterogeneous systems through additional keywords,
generating multiple versions of work unit kernels, and agglomerating
work units when required.  We will build on \cello's existing
capabilities by enhancing its \code{Field}, \code{Particle}, and other
\code{Data} objects to support multiple heterogeneous memory spaces,
and, working with the \charm\ group, we will develop a dynamic load balancing
method that maintains high data locality to improve the efficiency of
work unit agglomeration.  Together, \cello\ and \charm\ will help
isolate the complexity of heterogeneous hardware from \enzoe\ as much
as is feasible, while still providing a means for \enzoe\ numerical
methods to efficiently use multiple accelerators when available.

\section*{Acknowledgments}

We thank Sanjay Kal\'{e} and the Parallel Programming Laboratory at
the University of Illinois, Urbana Champaign for their helpful
discussions concerning using Charm++, and we thank John Wise for the
``extreme scales'' image in Fig.\ref{f:scales}.

\newpage
\bibliographystyle{IEEEtran}
\bibliography{IEEEabrv,biblio}


\end{document}

%% file: include.tex
\newcommand{\devel}[1]{}

\newcounter{figctr}

\usepackage{wasysym}
\usepackage{epsfig}
\usepackage{url}

\newcommand{\cello}{\textsf{Cello}}
\newcommand{\enzo}{\textsf{ENZO}}

\newcommand{\enzop}{\textsf{\mbox{Enzo-P}}}
\newcommand{\enzoe}{\textsf{\mbox{Enzo-E}}}

\newcommand{\charm}{\textsf{Charm\pp}}
\newcommand{\namd}{\textsf{NAMD}}
\newcommand{\openatom}{\textsf{OpenAtom}}
\newcommand{\changa}{\textsf{ChaNGa}}

\newcommand{\episimdemics}{\textsf{EpiSimdemics}}

\newcommand{\code}[1]{\textsf{#1}}
 
\newcommand{\pargraph}[1]{\devel{\P\ \textbf{#1} \\}}

\newcommand{\todo}{\devel{\Huge$\circ$} }
